\documentclass[aps,prl,reprint,groupedaddress,superscriptaddress,nofootinbib]{revtex4-1}
\usepackage[export]{adjustbox}
\usepackage{graphicx}% Include figure files
\usepackage{dcolumn}% Align table columns on decimal point
    \usepackage{bm}% bold math
\usepackage[utf8]{inputenc}
\usepackage{amsmath}
\usepackage{amsfonts}
\usepackage{amssymb}
\usepackage{graphicx}
\usepackage{color}
\usepackage{ bbold }
\usepackage{blindtext}
\newcommand{\bra}{\left\langle }
\newcommand{\ket}{\right\rangle  }
\usepackage{color}
\usepackage{ textcomp }
\usepackage[table,xcdraw]{xcolor}

\newcommand{\PT}{\mathbb{PT}}

\begin{document}

\title{Minimal constraints for Maximum Caliber analysis of dissipative steady state systems}
\author{Luca Agozzino}
\affiliation{Laufer Center for Physical and Quantitative Biology, Stony Brook University }
\affiliation{Department of Physics and Astronomy, Stony Brook University}
\author{Ken Dill}
\affiliation{Laufer Center for Physical and Quantitative Biology, Stony Brook University }
\affiliation{Department of Physics and Astronomy, Stony Brook University}
\affiliation{Department of Chemistry, Stony Brook University}

\begin{abstract}
\textit{Maximum Caliber} (Max Cal) is purported to be a general variational principle for Non-Equilibrium Statistical Physics (NESP).  But recently, Jack and Evans and Maes have raised concerns about how Max Cal handles dissipative processes. Here, we show that the problem does not lie in Max Cal; the problem is in the use of insufficient constraints.  We also present an exactly solvable single-particle model of dissipation, valid far from equilibrium, and its solution by Maximum Caliber.  The model illustrates how the influx and efflux of work and heat into a flowing system alters the distribution of trajectories. Maximum Caliber is a viable principle for dissipative systems.  
\end{abstract}

\maketitle

\section{The principle of Maximum Caliber for nonequilibrium processes}
Since the seminal work of Clausius and Boltzmann in the nineteenth century, predicting material equilibria has been based on the concept of entropy and its maximization. There has been a search for a more general variational principle that could also apply to nonequilibria, especially in the far-from-equilibrium regime. A good candidate has been the Principle of Maximum Caliber (Max Cal) \cite{jaynes1979maximum,jaynes1985complex,haken1986new,dewar2005maximum,presse2013principles,dixit2018perspective}, which is a Maximum-Entropy-like principle for inferring distributions over pathways and rate distributions of kinetic processes. Recently, concerns have been raised about whether Maximum Caliber handles dissipation properly.  We address those here, and show that Max Cal can handle dissipation properly when given appropriate constraints.

 Maximum Caliber is a method of inference about probability distributions over \textit{pathways} or \textit{trajectories}, in contrast to Maximum Entropy which infers distributions over \textit{microstates}.  Max Cal begins with a model of the accessible trajectories, $X=\{\xi(t_0),\xi(t_1),\xi(t_2)...\}$ of values $\xi$ at different times $t$. Max Cal infers the probability $p(X)$ of observing trajectory $X$ within trajectory space $\{X\}$ by maximizing the path entropy

\begin{equation}
\mathcal{S}=-\sum_X p(X)\log \frac{p(X)}{g(X)},
\end{equation}
 where the function $g(X)$ is some reference/prior distribution in the absence of constraints.  Now, in the simple situation of non-dissipative dynamics of a single dynamical quantity $J(X)$, for which the average,
 
\begin{equation}
\bra J\ket =  \int dX p(X) J(X)
\label{eq:meanJ}
\end{equation}
is known, the trajectory populations are obtained using the method of Lagrange Multipliers \cite{presse2013principles,dixit2018perspective}.

\section{Dissipative dynamics requires more constraints}

Two recent papers \cite{jack2016absence,maes2018non} assert that Max Cal will fail in some cases.  Jack and Evans (JE) \citep{jack2016absence} show that applying Max Cal with a single constraint to dissipative systems leads to the apparently inconsistent result of having no dissipation; Maes (M) \citep{maes2018non} asserts that problems whenever Max Cal is applied in cases of a time-symmetric component in one of the constraints. Here we clarify that these are not problems of the Principle of Maximum Caliber; these are problems of application of incomplete or incorrect constraints.

 We first address the JE situation.  Consider a dissipative system where a current $J(X)$ flows in conjunction with some finite amount of work $\delta w(X)$ done on the system and a heat flow $\delta q(X)$ out.  Assume that the statistical ensemble of all trajectories contains also those trajectories that are related to each other through a time-reversal transformation $\mathbb{T}$ and a space-reflection transformation $\mathbb{P}$ (it can also refer to reflection along only one of the physical coordinates \citep{jack2016absence}). For the right choice of current-generating force, the resulting current will always be antisymmetric under both time reversal and space-reflection transformations, so we assume that the forces acting on the system are of this type (an example is a shear stress, which generates a current with such property). As a consequence, under a combined $\mathbb{PT}$ transformation, the current will be identical to the untransformed current \cite{jack2016absence}. 
 
 Now consider the exchange of heat and work between the system and the external bath.  This will be antisymmetric under time reversal.  Running time backwards would reverse all three: the flow, the work, and the heat along the trajectory.  But, it will be invariant under space reflection.  No matter whether a force drives a current in a forward or backward direction along a trajectory, an identical amount of heat will be dissipated. This is not an assertion of reversibility of heat transfer; that would violate the Second Law of Thermodynamics. We are considering here only a single non-equilibrium trajectory, not a Second Law average over all trajectories.  Rather, it just means that if a trajectory has heat flowing into the system, its time-reversed trajectory has heat flowing out. As an example, consider a particle with mass $m$ sliding on a surface with friction coefficient $\phi$ and initial velocity $v$. The total energy dissipated through the process of slowing down until stopping is equal to the total kinetic energy $E_k=1/2 mv^2$ of the particle, which will increase the temperature of the surface by $\Delta T=E_k/C$, where $C$ is the surface's heat capacity. The time-reversed process would be the following: heating up the surface by exactly $\Delta T$ and wait for the thermal energy to spontaneously transform back into kinetic energy, accelerating the particle back to velocity $v$. This reverse process is extremely unlikely.  We illustrate a calculation of this probability in Max Cal below.

In general, for a dissipative system, a trajectory $X$ will have some current flow $J(X)$, at the same time as work $\delta w(X)$ performed on it, some heat dissipation $\delta q(X)$ out of it.  In this case, the $\PT$-reversed trajectory, $\mathbb{PT}X$, would have heat $\delta q(\mathbb{PT}X)=-\delta q(X)$ going into the system and work $\delta w(\mathbb{PT}X)=-\delta w(X)$ done on the external environment, because a space reflection transformation does not change the heat/work flow, but time reversal does. The probability of the transformed trajectory $\mathbb{PT}X$ should be much lower than of the untransformed trajectory $X$ for macroscopic currents, although we know from fluctuation theorems that for very small currents they can become comparable \cite{jarzynski1997nonequilibrium,crooks1999entropy}.  The result below agrees with such predictions.

For a dissipative steady state (DSS) the internal energy is unchanging with time, $\Delta U = \delta w+ \delta q$, because in the steady state, the heat out must equal the work in\footnote{Note, our convention is that energy going into the system is defined as positive.}.

 The argument of Jack and Evans is straightforward \cite{jack2016absence}.  First, they correctly note that if the only constraint is on $\langle J(X) \rangle$, (eq. \ref{eq:meanJ}), then maximizing the Caliber (i.e. the path entropy subject to the constraint) gives the following probability of trajectory $X:$

\begin{equation}
p(X)=\dfrac{e^{\mu J(X)}}{Z(\mu)},
\label{eqJE}
\end{equation}
where $Z(\mu) = \sum e^{\mu J(X)}$ is the sum of weights over all paths.  Second, since the flux is $\PT$ invariant, substitution of $J(X)= J(\PT X)$ into Eq~\ref{eqJE} gives the result that the probabilities must be $\PT$ invariant,

\begin{equation}
p(\PT X)=p(X).
\end{equation}

JE argue that such systems are not dissipative, because $\bra \delta q \ket = 0$, which they show as follows:

\begin{eqnarray}
&\bra \delta q \ket & =\int dX p(X) \delta q(X)\nonumber \\
&=&1/2 \int dX \left(p(X) \delta q(X)+  p(\PT X) \delta q(\PT X)\right)\nonumber \\
&=&1/2 \int dX p(X)(\delta q(X)-\delta q(X))=0,
\label{eq:diss}
\end{eqnarray}where the second line is obtained by considering that the Jacobian of a $\PT$ transformation equals 1.

JE conclude from this that Maximum Caliber cannot handle systems, such as a sheared fluid, that are dissipative.  On the contrary, we show below that the problem above is the use of only a single constraint, namely $\langle J(X) \rangle$.  This misses the essentiality of the coupling of the flow $J$ inside the system to the work and heat flows into and out of the system.  The latter require additional constraints. 

\section{The number of constraints must at least equal the number of independent flow variables}

To illuminate the problem, consider the corresponding situation in equilibrium thermodynamics.  The equilibrium entropy can be expressed as a function $S = S(U, V, N)$ of three independent extensive variables -- energy, volume and particle number.  If all three independent variables are free to change in a process, you cannot adequately specify the state of the system with only a single Lagrange multiplier, say the pressure $p$; you must also specify the temperature $T$ and chemical potential $\mu$.  You need a Lagrange multiplier for every independent variable. 

 In dissipative dynamical systems too, there are multiple independent variables.  You can specify an average flow rate $\langle J \rangle$, but dissipative systems also entail heat and work flows in and out, and those can affect the trajectory distribution.  For example, you can achieve a given average particle flow rate in multiple ways, such as increasing the work done on the particle in a medium of increasing viscosity that dissipates more heat.  Predicting the trajectory distribution in dissipative systems requires knowing the heat and work rates, not just the particle flow rate.\footnote{Note, however, that while $(U,V,N)$ are conserved quantities in the equilibrium metaphor, $J_q$ and $J_w$ are not necessarily conserved in flow situations.}

 For example, consider particles flowing along the axis of a tube, with an average current of $\langle J(X)\rangle=J$. That particle flow can be independent of the rate of work flow $\bra J_w(X)\ket$ and heat flow $\bra J_q(X)\ket$ into and out of the tube.  Some situations will reduce these 3 variables to fewer; other situations will not.
 
First, consider any steady-state flow, dissipative or not.  By definition, the total internal energy will be unchanging with time, $\Delta U=0$.  So, it follows from the First Law that

\begin{equation}
\delta q=-\delta w.
\label{eq:QW}
\end{equation}
Thus, in steady-state flows, the heat current must equal the work current,
\begin{equation}
\bra J_q(X)\ket=-\bra J_w(X)\ket
\end{equation}
where our convention is that current flows into the system are defined as positive.

Now, in a non-dissipative steady state (nDSS), we have $\bra J_q(X)\ket=-\bra J_w(X)\ket = 0$, leaving us only one independent variable, $J$.  However, in a dissipative steady state (DSS), energy must continuously enter the system in order to sustain the current $J$, so now we have 3 constraints,

\begin{eqnarray}
&\langle J(X)\rangle&=J\\
&\langle J_w(X)+J_q(X)\rangle&=0\\
&\frac{1}{2}\langle J_w(X)-J_q(X)\rangle&=J_\text{E}
\label{constraints3}
\end{eqnarray}where $J_\text{E}$ is the energy influx rate.

 We note two points here.  First, if were to use only a single constraint $\langle J \rangle$ for a DSS, as in the JE argument, it is tantamount to setting $J_E = 0$ above, thus effectively asserting that heat and work flow in and out are both zero, and thus that the system is, by definition, not dissipative.  Second, near equilibrium and for non-steady states, dissipation $\bra J_q \ket$ is proportional to the current $\bra J\ket$, so in that case a single constraint can be sufficient to describe the system \cite{hazoglou2015communication}. 
 
 Therefore, for steady states, arbitrarily far from equilibrium, only non-dissipative systems can be described when only a single constraint, $\langle J \rangle$, is specified.

\section{For dissipative steady-states, Maximum Caliber requires at least 3 constraints.}

For DSS situations, with constraint eqs~\ref{constraints3}) above, the expression for Caliber is:

\begin{eqnarray}
\mathcal{C}=-\int dX p(X)\ln p(X)\\
-\alpha \left( \int dX p(X)-1\right)\\
-\mu \left( \int dX p(X)J(X)-J\right)\\
-\nu \left( \int dX p(X)J_w(X)-J_\text{E}\right)\\
-\lambda \left( \int dX p(X)J_q(X)+J_\text{E}\right)
\label{eq:caliber}
\end{eqnarray}
where we chose here, for simplicity, to define each current individually instead of constraining the sum and the difference. Maximizing Caliber gives the trajectory probabilities as

\begin{equation}
p(X)=\dfrac{e^{\mu J(X)+\nu J_w(X)+\lambda J_q(X)}}{Z(\mu,\nu,\lambda)}
\end{equation}
where $Z =\sum e^{\mu J(X)+\nu J_w(X)+\lambda J_q(X)}$.

This Max Cal formulation shows that reverse trajectories in dissipative processes are unlikely for large currents.  Using the $\PT$ transformation, we can calculate the relative probability that a system would absorb heat from the environment (and produce work):

\begin{equation}
\dfrac{p(\PT X)}{p(X)}=e^{-2(\nu J_w(X)+\lambda J_q(X))}.
\label{eq:ratio}
\end{equation}
This fluctuation relation shows that `wrong-way' paths, which take up heat in dissipative flows, become exponentially improbable with increasing current, as they should.  If the only constraint here were on $\langle J \rangle$, as in JE, then $\bra J_q\ket=\bra J_w\ket=0$ and wrong-way flows would be predicted to be much more probable.

The Max Cal procedure gives the distribution of all the trajectories. On the one hand, it uses as an input constraint, the heat uptake $J_q(X)$ averaged over all the trajectories:
\begin{equation}
\bra\delta q\ket= \bra J_q\ket \Delta t =\Delta t \int dX p(X)J_q(X).
\end{equation}
On the other hand, Max Cal then gives as a prediction the higher moments, such as the mean-square fluctuations of the heat: 

\begin{equation}
\bra  \delta q^2\ket =\Delta t^2 \int dX J_q^2(X) p(X).
\end{equation}

\section{A solvable model of a dissipative system: a particle in 1-dimensional flow, with heat and work.}

\begin{figure}
\includegraphics[width=.8\linewidth]{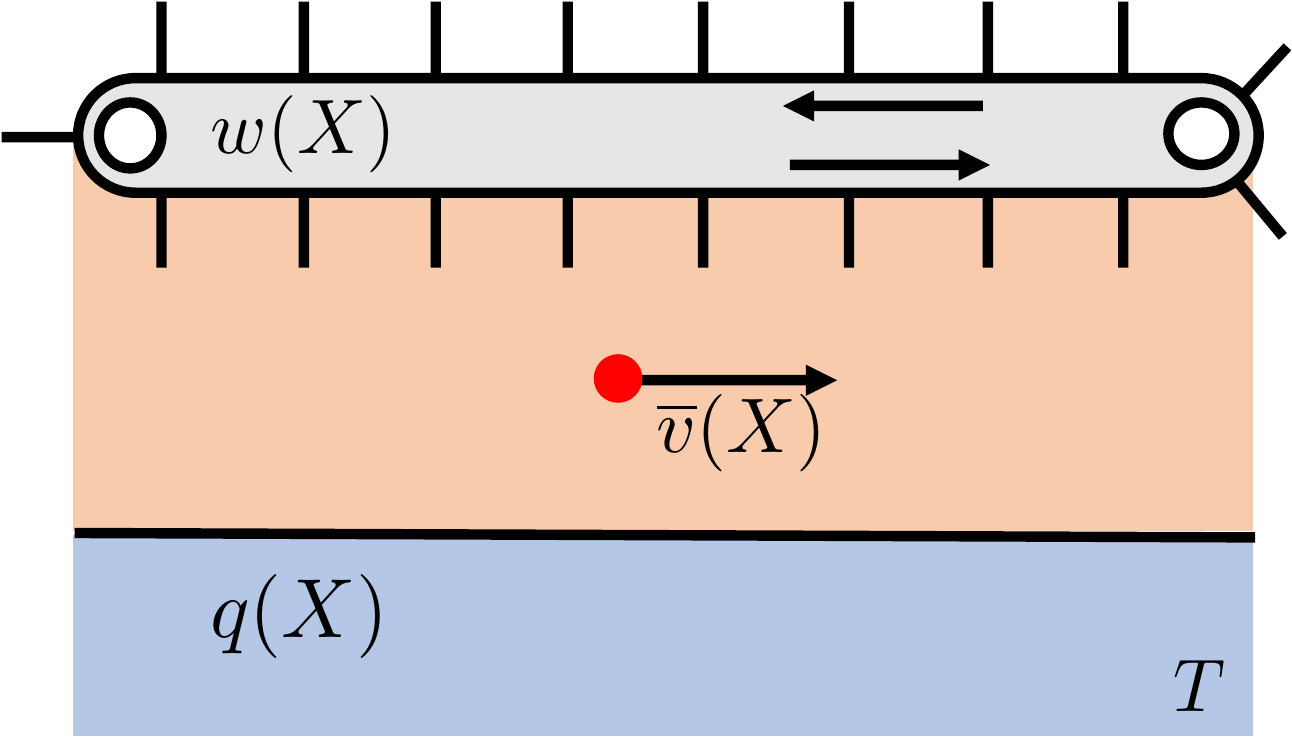}
\caption{\textbf{A particle in a dissipative system.} The particle can receive energy from the belt or from the thermal bath, but it can also transmit energy to the belt by hitting it or to the thermal bath, by friction on the walls of the conduit.}
\label{fig:particle}
\end{figure}

In this section, we illustrate with a concrete model.  Consider one particle moving inside a 1D conduit. The particle is in contact with an external thermal bath with which it can exchange heat.  The particle can also interact with a conveyor belt that performs work from outside to boost the particle's velocity; see Fig. \ref{fig:particle}.

A trajectory $X$ is a series of $N$ steps, each one of which takes time $\Delta t$.  In each time step, the particle experiences one of three possibilities: (i) it increases or decreases its velocity by $\Delta v$, by collision with the belt, (ii) it increases or decreases its velocity by $\Delta u$ by exchanging heat with the bath, or (iii) it undergoes no change in velocity in that time step.  A full trajectory is a string of such states: up, up, stay, up, down, up, .... for example. The quantities $\Delta v$ and $\Delta u$ are not limited to a fixed value, but can be anything within a given range.

The trajectory for a given particle has three identifying quantities. The average velocity of the particle along the trajectory $\overline{v}(X)$, the work done on the particle by the belt $w(X)$ and the heat absorbed by the particle from the thermal bath $q(X)$. As a convenient convention, we take both $w(X)$ and $q(X)$ to be positive when the energy flows from the external environment to the particle, so for the work, this convention is the opposite with respect to the one used in thermodynamics. Note that for the average velocity of a given trajectory $\overline{v}(X)$ we have used the overbar symbol to distinguish it from a trajectory-ensemble average; $\overline{v}(X)$ is just the average velocity maintained by the particle in a specific trajectory, whereas we would use the symbol $\bra\overline{v}(X)\ket\equiv \sum p(X)\overline{v}(X)$ to refer to the trajectory-ensemble average, hence averaged over all the possible trajectories.

This allows us to enforce some minimal constraints which identify a DSS without ambiguity. The constraints are the following:

\begin{eqnarray}
&\bra w(X)\ket &= E_\text{in}\label{eq:constr1}\\
&\bra q(X)\ket &= -E_\text{in}\\
&\bra \overline{v}(X)\ket &= V\label{eq:constr3}
\end{eqnarray}where $E_\text{in}$ is the average work input (or negative heat output).

The particle starts at time $t=0$ with velocity $v_0$. So, a given trajectory $X$ can be specified by an initial velocity and a sequence of changes in velocities:

\begin{equation}
X=\lbrace v_0, \xi_1, \xi_2,..., \xi_{N-1}\rbrace
\end{equation}
where $ \xi_j=\Delta v_j$ or $\Delta u_j$, where $j$ is an index of the time step, depending on which processes occurred along the given trajectory. Now, Maximum Caliber gives the probability of a given DSS trajectory as

\begin{equation}
p(X)= p(v_0, \xi_1, \xi_2,..., \xi_{N-1}) = \dfrac{e^{\nu w(X)+\lambda q(X)+ \mu \overline{v}(X)}}{Z}
\label{eq:probability}
\end{equation}

All the functions $w(X)$, $q(X)$ and $\overline{v}(X)$ can be expressed in terms of the particular sequence of velocity changes in trajectory $X$.  

Now, under a $\PT$ transformation, each trajectory function is transformed as follows:

\begin{eqnarray}
&w(\PT X) &= -w(X)\\
& q(\PT X) &= -q(X)\\
&\overline{v}(\PT X) &= \overline{v}(X).
\end{eqnarray}
This is because both heat and work are invariant under space reflection, but are anti-symmetric under time reversal. Therefore, the ratio between the $\PT$-transformed and untransformed trajectory is

\begin{equation}
\dfrac{p(\PT X)}{p(X)}=e^{-2[\nu w(X)+\lambda q(X)]}
\label{eq:ratio2}
\end{equation}
which does not equal 1, except in the non-dissipative case that the trajectory does not involve any energy exchange\footnote{In this case, the $\PT$-reversed trajectory must have identical probability, because it is the identical trajectory.}.

For a general $N$-step process, the functional form is too complex for analytical solution, due to the non-linear relation between velocity and kinetic energy: the change in velocity at time step $n$ will depend upon all the changes in velocity at time steps $n-1$, $n-2$, ..., $0$.

The partition function can be calculated numerically in that case, and the values of the Lagrange multipliers can be tuned to make sure that constraint averages are satisfied. In the next section we will show how to solve the problem analytically in an even simpler case.

\subsection{Simplified trajectories with only 3 time steps}
Now, we can obtain a closed-form expression if we further simplify the model above to just 3 total time steps.  Any trajectory is now described by the vector
\begin{equation}
X=\{v_0,\xi_1,\xi_2\}
\label{3step}
\end{equation}
 where $v_0$ is the initial velocity of the particle (first step), $\xi_1$ is the change in velocity in the second step and $\xi_2$ is the change in velocity in the third step. At steps 2 and 3, the velocity can either remain the same ($\xi_i=0$) or change by interaction with the moving belt ($\xi_i=\Delta v_i$) or change by heat exchange ($\xi_i=\Delta u_i$) (see Supporting Material for details). Again, the functional form of the probability is given by Eq. \ref{eq:probability}, but now just for the short-trajectories of Eq. \ref{3step}.

The Max Cal dynamical partition function is obtained by computing the following sum over all the small number of trajectories $X$:

\begin{equation}
Z=\sum_X e^{\nu w(X)+\lambda q(X)+ \mu \overline{v}(X)}
\label{eq:partition}
\end{equation}
In order to correctly express the form of the sum in Eq.~\ref{eq:partition} we take into account the fact that at every step we are assuming that only one type of velocity change is possible, either heat driven or work driven.

We can compute the change in particle velocity that is due to work or heat exchange.  In the Supporting Material, we calculate the sum in Eq.~\ref{eq:partition} and solve Eqs.~\ref{eq:constr1}-\ref{eq:constr3}, to obtain the following values of the Lagrange multipliers:

\begin{eqnarray}
\mu &\simeq&\frac{3\eta}{V}\\
\nu &\simeq& \frac{\epsilon}{2E_\text{in}}\\
\lambda &\simeq&-\frac{\epsilon}{2E_\text{in}}
\end{eqnarray}
where $\eta=V^2/V_\text{max}^2$ and $\epsilon=\Delta V_Q/\Delta V_W$. $V_\text{max}$ is the maximum velocity that the conduit can withstand, $\Delta V_Q$ is the maximum change in velocity due to heat exchange and $\Delta V_W$ the one due to heat exchange.

In order to obtain this result, we have assumed that the measured velocity $V$ is much smaller than the maximum rate $V_\text{max}$, so $\eta<<1$.  We also assumed that the maximum change in velocity due to work is much larger than the one due to heat, because work is always directed in a specific direction, so this means $\epsilon<<1$. Such assumptions, although not necessary to solve the problem, make it easier to obtain an analytical expression for the Lagrange multipliers.  

The trajectory probabilities in this 3-step model are:

\begin{equation}
p(X)=\frac{1}{Z}\exp\left\lbrace\epsilon\frac{w(X)-q(X)}{2E_\text{in}}+3\eta\frac{\overline{v}(X)}{V}\right\rbrace
\label{3p}
\end{equation}

\begin{figure}
\includegraphics[width=\linewidth]{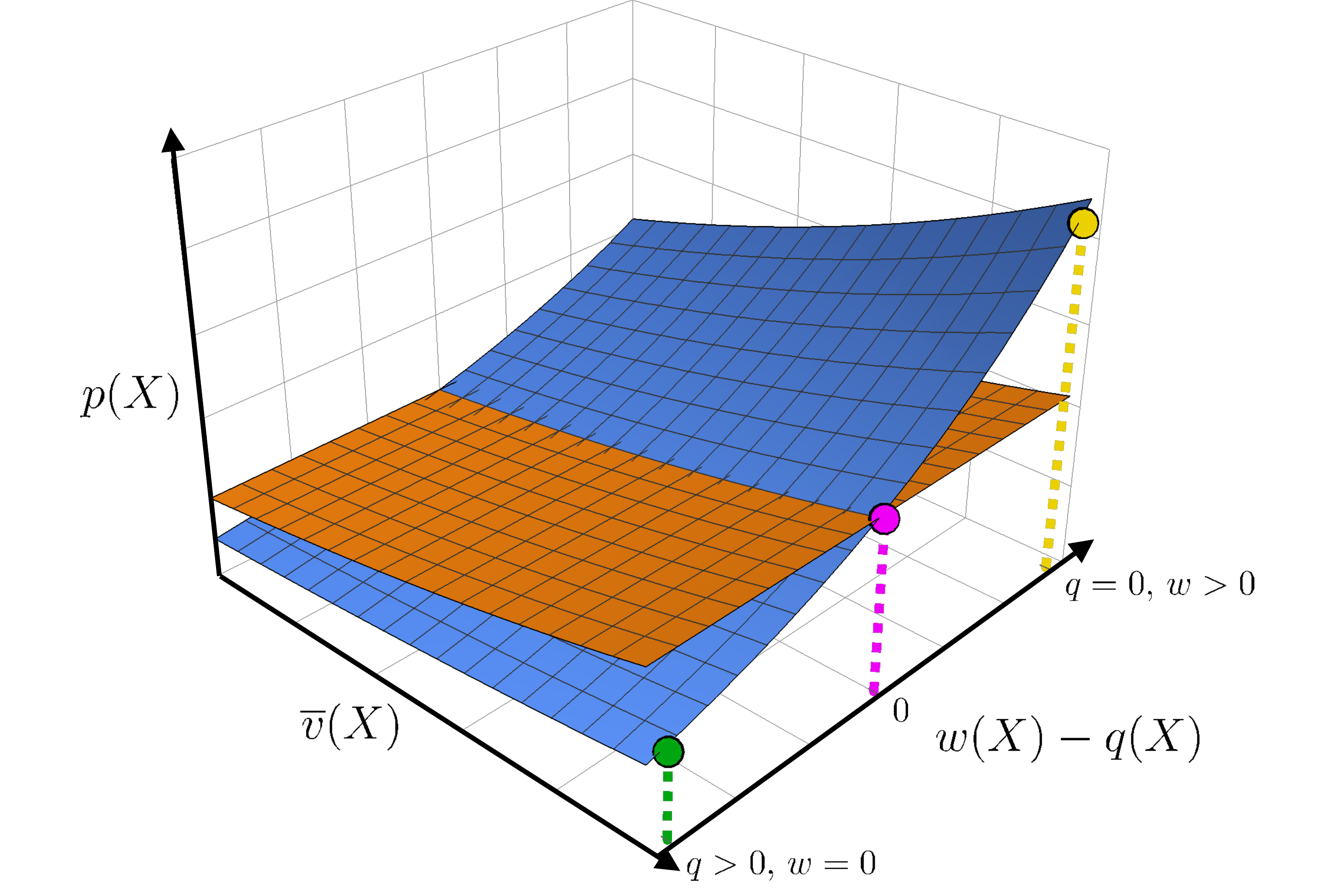}
\caption{\textbf{The Max Cal probability distribution vs JE.} Using 3 constraints (blue plane) it is able to capture the difference between trajectories with different energy sources, which is not possible when only one constraint is used (orange plane). The coloured dots show the difference of the probability of three trajectory with the same average velocity but different energy source, as depicted in Fig.~\ref{fig:3-Trajectories}.}
\label{fig:3dPlot}
\end{figure}

Eq \ref{3p} computes the probability of any pathway $X$ for fixed values of the two observables, $E_\text{in}$ and $V$. Fig \ref {fig:3dPlot} shows an example of trajectory populations as a function of the three properties, $\overline{v}(X)$, $w(X)$ and $q(X)$ of each trajectory, for fixed values of $V$, the particle flow velocity, and for fixed energy input $E_\text{in}$. The orange plane in Fig \ref {fig:3dPlot} shows what you would predict if you knew only the mean flow velocity $\overline{v}(X)$.  The trajectory population would not depend on the source of energy into the system (heat or work).  The blue plane shows two things: (1) how trajectories that have a higher speed and in the same direction of the average $V$ become more populated, and (2) the trajectories become more populated when more work flows in and heat is dissipated ($w(X)-q(X)>0$), and become less populated as more energy flows in to the system from the external bath, producing work ($w(X)-q(X)<0$).

From Eq \ref{3p}, we can readily compute the ratio of probabilities for $\PT$-reversal:
\begin{equation}
\dfrac{p(\PT X)}{p(X)}\simeq e^{-2\epsilon\dfrac{w(X)-q(X)}{E_\text{in}}}
\label{eq:ratio3}
\end{equation}
Eq \ref{eq:ratio3} shows that for a given amount of energy that is put into the system, a trajectory that has a large dissipative current is more likely than the $\PT$-reversed, non-dissipative one. Eq \ref{eq:ratio3} also correctly predicts that when energy exchange is small, the probability of a wrong-way flow is comparable to a right-way flow.

\begin{figure}[h]
\includegraphics[width=.9\linewidth]{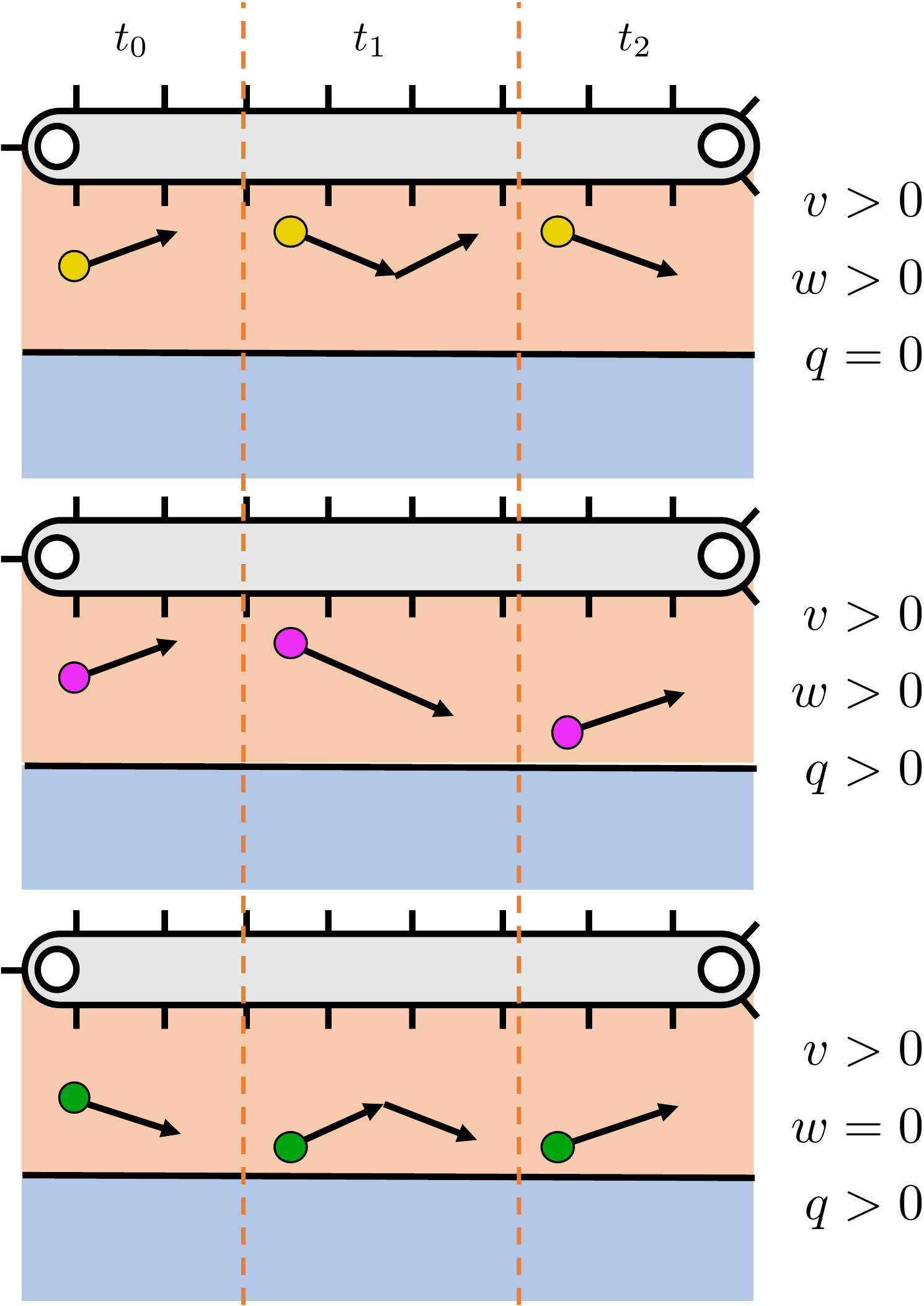}
\caption{\textbf{Three possible trajectories for a single particle}. \textbf{Top:} The particle interacts in both steps with the conveyor belt, receiving energy as work; \textbf{Center:} The particle first receives work then heat from the thermal bath; \textbf{Bottom:} The particle only receives energy as heat from the thermal bath. In Fig.~\ref{fig:3dPlot} the corresponding coloured dots with the respective probabilities.}
\label{fig:3-Trajectories}
\end{figure}

In this way, Max Cal captures the difference between trajectories having the same average velocity but caused by very different processes. In Fig~\ref{fig:3-Trajectories} we show three examples of trajectories, each with the same average velocity $v$ but with different values of $w$ and $q$. The first trajectory corresponds to the process in which the particle is hit twice by the belt; in the second the particle is hit first by the belt and then it receives energy from the thermal bath; in the third, the particle receives energy from the bath twice. If the process were non-dissipative, the three trajectories would have the same probability, but in this dissipative case, Max Cal shows how the probabilities are different (Fig~\ref{fig:3dPlot}).

\section{The Maes Argument and the proper number of constraints}
Maes argument \citep{maes2018non} is a bit more subtle, because it points out the fact that when the only chosen constraints are time-asymmetric currents, the only possible outcome is a system without any dissipation. We agree in principle that this is the case, but this would be a problem of making a poor choice of constraints, and not with Max Cal itself.  When there is available knowledge of the system that is being ignored, like work or heat transfer, (or in general what Maes calls a \textit{frenetic} contribution \cite{maes2018non}), it is to be expected that Max Cal will not necessarily be consistent with it. In this case too, the problem is with the choice of constraints, not the Max Cal principle.  One further point is that our discussion here considers only the 3 constraints needed for DSS; some different situations may need more or different constraints.

\section{Conclusions}

We have shown here that Maximum Caliber can handle dissipation properly, but it requires the application of appropriate restraints.  In DSS, you need both the mean rate of flow,and also the work performed on the system and the heat that is dissipated. We show this on general grounds, but we also give a specific solvable model of a single particle flow that is subjected to heat and work input and output. This toy model may be useful for studying dissipative flows.

\section{Acknowledgments}
This research was supported by the National Science Foundation grants PHY1205881 and MCB1344230 and the Laufer Center for Physical and Quantitative Biology. We wish to thank Jason Wagoner, Purushottam Dixit and Kingshuk Ghosh for the precious insights and discussions.

%\pagebreak

\bibliography{bibliography}

\end{document}